\newcommand{\nuc}[2]{$^{#1}$#2}
\newcommand{\nucm}[2]{$^{#1}$#2$^m$}

\newcommand{\qec}{$Q_{EC}$}

\def\F{{\cal F}}

\documentclass[twocolumn,showpacs,preprintnumbers,amsmath,amssymb]{revtex4}
\usepackage{epsfig}
\usepackage{dcolumn}% Align table columns on decimal point
\usepackage{bm}% bold math
\begin{document}

\preprint{APS/123-QED}

\title{$Q$-values of the Superallowed $\beta$-Emitters \nucm{26}{Al}, \nuc{42}{Sc} and \nuc{46}{V} \\
and their impact on $V_{ud}$ and the Unitarity of the CKM Matrix }
% Force line breaks with \\

\author{T. Eronen} \email{tommi.eronen@phys.jyu.fi}
\author{V. Elomaa}
\author{U. Hager}
\author{J. Hakala}
\author{A. Jokinen}
\author{A. Kankainen}
\author{I. Moore}
\author{H. Penttil\"{a}}
\author{S. Rahaman}
\author{J. Rissanen}
\author{A. Saastamoinen}
\author{T. Sonoda}
\author{J. \"{A}yst\"{o}}
\affiliation{Department of Physics, P.O. Box 35 (YFL), FIN-40014 University of Jyv\"{a}skyl\"{a}, Finland}
\author{J.C. Hardy}
\affiliation{Cyclotron Institute, Texas A\&M University, College Station, Texas 77843}
\author{V. S. Kolhinen}
\affiliation{Sektion Physik, University of Munich (LMU) Am Coulombwall 1, 
D-85748 Garching, Germany}

\date{\today}% It is always \today, today,

\begin{abstract}
The $\beta$-decay $Q_{EC}$~values of the superallowed beta emitters
$^{26}$Al$^m$, $^{42}$Sc and $^{46}$V have been measured with a Penning
trap to a relative precision of better than $8\times 10^{-9}$. Our result for $^{46}$V,
7052.72(31)~keV, confirms a recent measurement that differed significantly from
the previously accepted reaction-based $Q_{EC}$~value.  However, our
results for $^{26}$Al$^m$ and $^{42}$Sc, 4232.83(13)~keV and 6426.13(21)~keV, are
consistent with previous reaction-based values.  By eliminating the possibility of a
systematic difference between the two techniques, this result demonstrates that no
significant shift in the deduced value of $V_{ud}$ should be anticipated.

\end{abstract}

\maketitle

A recent critical survey of superallowed $0^+\rightarrow 0^+$ nuclear $\beta$-decays \cite{Ha05a}
presented a remarkably consistent picture, from which it was possible to obtain precise values
and demanding limits on a number of fundamental weak-interaction parameters \cite{Ha05a,Ha05b}.
In particular, the value of the up-down element of the Cabibbo-Kobayashi-Maskawa (CKM) matrix was
determined from the superallowed data to be $V_{ud}$ = 0.9738(4).  Since $V_{ud}$ is a
key component of the most demanding available test of the unitarity of the CKM matrix, the
precision and reliability of the value for $V_{ud}$ has a direct impact on the search for physics
beyond the standard model.

Shortly after the survey was published, a new \qec-value measurement was reported by Savard
{\it et al.} \cite{Sa05} for the superallowed $\beta$-decay of \nuc{46}{V}.  This was the first
time that the \qec~value for any of the nine most-precisely known superallowed transitions had been
measured with an on-line Penning trap.  All previous results had been obtained from reaction $Q$
values: generally from (p,n) or ($^3$He,t) reactions, or from a combination of (p,$\gamma$) and
(n,$\gamma$) reactions.  Stunningly, the Penning-trap result differed significantly from the survey
result and left the $\F t$ value for the \nuc{46}{V} transition anomalously high with respect to the 
$\F t$ values for the other superallowed transitions.  There was understandable concern that this
could be signaling a previously undetected systematic error in all the reaction measurements, which, when
corrected, might lead to a significant shift in $V_{ud}$ from the value obtained in the survey.

Since systematic changes of only a few hundred eV in the \qec~values could have an appreciable effect
on $V_{ud}$, this concern prompted a careful study \cite{Ha06} of whether such systematic errors
could be excluded in past measurements of (p,$\gamma$) and (n,$\gamma$) reaction $Q$ values.  The study's
authors concluded that systematic effects up to at least 200 eV could not be excluded, and they proposed that
a Penning-trap measurement of the superallowed transition from \nucm{26}{Al} would provide an excellent
case to test for systematics since the corresponding reaction-based $Q$ value was particularly
soundly based.

We report here Penning-trap measurements of the \qec~values for three superallowed $\beta$ transitions.
The first, the decay of \nuc{46}{V}, was chosen to confirm (or not) the recent unexpected Penning-trap
result \cite{Sa05}.  The second, \nucm{26}{Al}, is the case proposed \cite{Ha06} as a test for
systematic effects; and the third, \nuc{42}{Sc}, is another case in which high-quality (p,$\gamma$)
and (n,$\gamma$) reaction measurements have previously been performed.  The measurements were
specifically aimed at establishing whether undetected systematic effects were present in earlier
measurements and whether a significant change in $V_{ud}$ might be anticipated as a result. 

All ions of interest were produced at the IGISOL facility \cite{hui04}.  We produced
\nuc{46}{V} and $^{26}${Al}$^m$ via ($p$,$n$)-reactions,
with 20- and 15-MeV proton beams incident on enriched \nuc{46}{Ti} and \nuc{26}{Mg} targets respectively.
For $^{42}$Sc, we used a $^3$He beam of 20~MeV on $^\textrm {nat}$Ca. In these
bombardments, not only were the superallowed emitters of interest produced in the primary reactions 
but ions from the target material itself -- the $\beta$-decay daughters of these emitters -- were
also released by elastic scattering
of the cyclotron beam. The recoil ions were slowed down
and thermalized in the gas cell of an ion guide filled with 150~mbar of helium
\cite{hui04}. These were then transported by gas flow and electric fields
through a differentially pumped electrode system into a high-vacuum region, accelerated
to 30~keV and passed through a 55$^\circ$ dipole magnet for a coarse
mass selection with resolving power of 300-500.

The mass-separated ion beam was then transferred to the JYFLTRAP setup.
This consists, first, of a radio-frequency
quadrupole (RFQ) cooler \cite{nie01b}, which is used to  
improve the quality of the beam and bunch it for efficient injection into
the Penning-trap system.  The latter consists of two cylindrical
traps housed inside the same superconducting 7-T magnet. The first trap
is filled with helium buffer gas to allow for purification of the
ion sample. With successive magnetron dipole excitation and
mass-selective quadrupole excitation, a mass resolving power 
of up to a few$\times 10^5$ \cite{kol03} can be achieved in this
first trap, which is enough to resolve the isomeric and ground states
in \nuc{26}{Al} and \nuc{42}{Sc}.

\begin{figure} 
\epsfig{file=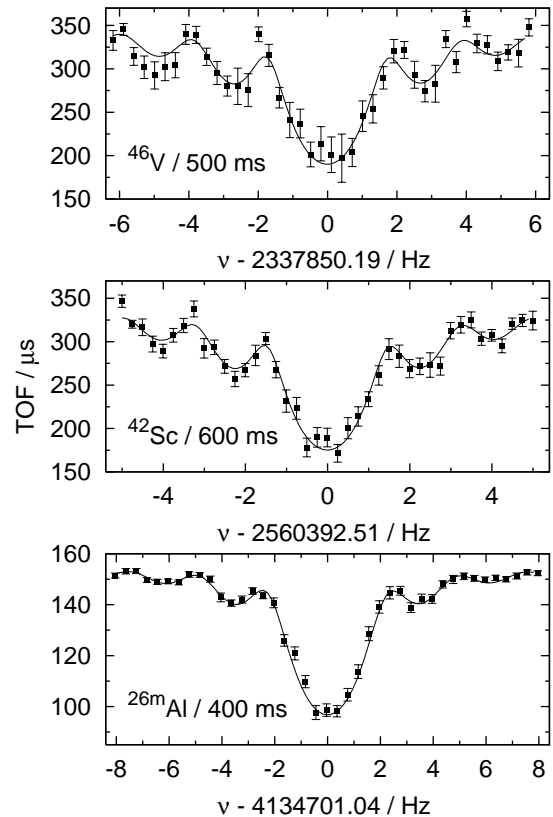,width=0.9\columnwidth}
\caption{\label{fig:resonances}Examples of the time-of-flight resonances measured for each of the superallowed
beta emitters.  The solid curves are fitted functions (see text). }
\end{figure}

After purification, the ion ensemble was injected into the second Penning trap for the actual
mass measurement. A dipole excitation was used to establish a magnetron orbit with
a fixed frequency and amplitude. Then, the ions were exposed to a radiofrequency quadrupole
electric field for a given time. The amplitude of the RF electric field was tuned
so that, when the frequency corresponded to the cyclotron frequency of the ion of
interest, the whole magnetron motion was converted to cyclotron motion.
After the quadrupole excitation, the ions were extracted from the trap and their time-of-flight
to the micro channel plate detector recorded. The frequency corresponding to the shortest time-of-flight
is the true cyclotron frequency \cite{blo53, kon95}. To locate the precise resonance frequency,
we scanned the frequency and recorded the time-of-flight over a range that spanned the resonance.
Examples of these frequency scans appear in Fig. \ref{fig:resonances}.

The \qec~value of each ion of interest was obtained directly from the frequency ratio
of the mother and the daughter nuclei. The cyclotron frequency measurements were interleaved: first
we recorded a frequency scan for the daughter, then for the mother, then for the daughter and so
on. This way, the slow drift of the magnetic field, mostly due to drifts in the room 
temperature, could be treated properly by interpolation of the reference frequency
to the instant of measurement for the ion of interest.  In the cases of \nucm{26}{Al}
and \nuc{42}{Sc} we also measured the resonance frequencies of the nearby high-spin, long-lived
states to check for consistency.

For each measurement, data were collected in several sets. Each set comprised $\sim$10 pairs
of parent-daughter frequency scans taken under the same conditions. Between sets, the excitation
time was changed. Each of the resonance curves was fitted with a realistic function, described
in Ref. \cite{kon95}, which yielded values for the resonant frequency and its statistical
uncertainty.

\begin{figure}[t]
\epsfig{file=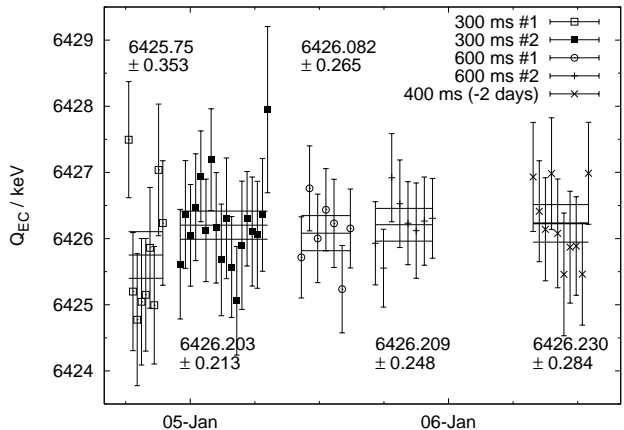,width=\columnwidth}
%\begin{figure} 
%\includegraphics{Sc-Ca}
\caption{\label{fig:setti}Individual \qec~values obtained for \nuc{42}{Sc} in January 2006.  Each
set of measurements corresponds to a different excitation time (see legend). The
uncertainties given for each set contain only statistical uncertainties from fitting and those
arising from short-term magnetic field fluctuations.}
\end{figure}

For \nuc{46}{V}, a total of 40 resonances were obtained with \nuc{46}{Ti} as a reference ion;
these were grouped in three sets with excitation times of 700, 500 and 300 ms.  For
\nuc{42}{Sc} we used \nuc{42}{Ca} as a reference and obtained 52~resonances in 5 different sets
covering three different excitation times, 300, 400 and 600 ms (see Fig. \ref{fig:setti}).
As a consistency check for \nuc{42}{Sc}, we also measured the \qec~value of \nucm{42}{Sc},
referenced both to \nuc{42}{Ca} and to \nuc{42}{Sc}.

The \nucm{26}{Al} measurement followed the same pattern as for \nuc{42}{Sc}. The resonances of
\nucm{26}{Al} and \nuc{26}{Al} were both measured with respect to the ground state of \nuc{26}{Mg}
and, in addition, we measured the excitation energy of \nucm{26}{Al} directly by using
\nuc{26}{Al}(gs) as a reference. In each of these three ratio measurements, excitation times
of 200, 300 and 400~ms were used. As a further consistency check, the frequency ratios for
\nucm{26}{Al} and \nuc{26}{Al} were also obtained with \nuc{25}{Mg} as the reference ion;
however, in this latter case only a 200-ms excitation time was used and relatively few resonances
were obtained.  Our final measured frequency ratios for all cases are given in table \ref{tab:results}. 

With the frequency ratios thus determined, we derived the \qec~value between mother-daughter
pairs from the following equation: 
\begin{equation} \label{eq:q_value}
Q_\mathrm{EC} = m_{m} - m_{d} = \left ( \frac{\nu_{d}}{\nu_{m}} -1 \right ) m_d,
\end{equation}
where $m_{m}$ and $m_{d}$ are the masses of the singly charged mother and
daughter ions and $\frac{\nu_{d}}{\nu_{m}}$ is their frequency ratio.
In our experiment, all measured ions were singly charged and the mass excess values
for the reference ions, $m_{d}$, were obtained from Ref. \cite{aud03}. Since
the term inside parenthesis is small ($<10^{-3}$), the uncertainty contribution from
$m_{d}$ to the \qec~value is negligible.  The final \qec~value (or, where appropriate, the
excitation energy, $E_\mathrm{ex}$) for each pair was obtained from the weighted average of the
results from each relevant set.  The results from successive sets of scans in our \nuc{42}{Sc}
measurement are shown in Fig. \ref{fig:setti}. 

In obtaining a final uncertainty on the frequency ratios (and derived \qec~values), we
considered more than just the statistical
uncertainties in the fitted resonance frequencies.  Although the slow, linear drift of the magnetic
field was accounted for by the interpolation process already described, short-term field
fluctuations may also exist, so we quadratically added an uncertainty of 47, 20 and 34 mHz 
to the individual frequency uncertainties for ions with $A=46$, 42, and 26, respectively. These
numbers were derived from the observed scattering in the reference frequencies, and have %only a
%Changed: only a -> relatively
relatively
small impact on the final uncertainties.

An important source of systematic uncertainty is the number of ions stored simultaneously
in the trap, which can cause shifts in the resonance frequency.  There are two ways to deal
with this effect.  The first is simply to keep the number of ions stored simultaneously
in the trap small. We took this approach in the cases of \nuc{46}{V} and in \nuc{42}{Sc}, for
which we filtered the data during analysis, only including time-of-flight results from bunches
that included 2 ions or less. We took account of any possible remaining systematic shift due
to this non-ideal countrate by including an additional uncertainty of $\sqrt{2}\times$0.008~Hz,
as was done in our earlier measurement of \nuc{62}{Ga} \cite{ero06}.

The second way of dealing with the count-rate effect is to divide the data into different groups
depending on the number of ions per bunch that were detected.  Each group is fitted separately
and a resonance frequency obtained.  The resonant frequencies obtained for the various numbers
of detected ions are then extrapolated back to 0.6~detected ions per bunch, a value that
corresponds to 1~ion stored per bunch corrected for the known detection efficiency of 60\%
\cite{kel03}.  Only in the case of \nucm{26}{Al} did we have high enough statistics to allow
us to analyze the data following this procedure.

Finally, because most of our measurements were of mass doublets, each partner having the same
mass number, mass-dependent systematic effects are expected to cancel out in these cases.  Only
in the case where \nuc{25}{Mg} was used as a reference for A=26 ions, was it necessary for us
to add an additional uncertainty of $1\times 10^{-8}$ to the derived results.

Our results for the \qec~(or $E_{ex}$) values 
are given in Table \ref{tab:results} for each measured doublet.  Also given are the derived
mass excesses for each identified ion.  Our data for both A=26 and A=42 allow us to obtain the
superallowed \qec~values by two routes: via the direct doublet measurement and via the combination of
the other two doublets involving the non-$0^+$ state in the mother nucleus.  In both cases
the two routes led to statistically consistent results, and it is their average that we quote
for our final \qec~values.

We note as well that for both A=26 and A=42, we obtain the excitation energies of the isomeric
states in the mother nuclei.  In the case of \nucm{26}{Al} we obtain this energy via three
different routes, giving an average result of 228.27(13) keV.  This compares very favorably
with the accepted value \cite{En90} of 228.305(13), which is based on $\gamma$-ray measurements.
For \nucm{42}{Sc} the two paths we have available yield an average excitation energy of
616.61(22), which also is in tolerable agreement with 616.28(6), the accepted value \cite{En90}.
These results provide a gratifying confirmation of the consistency of our measurements.

\begin{table*}
\begin{center}
\caption{\label{tab:results}Results of the present measurements. The symbol \# denotes the number of
doublet measurements made.  The superallowed decay branches are indicated by bold type. % The tabulated
%mass-excess values were derived from Eq \ref{eq:q_value} with 
The reference mass excesses were taken from
Ref. \cite{aud03}.}
\vskip 1mm
\begin{ruledtabular}
\begin{tabular}{llllll}	
ion 		& reference 	& \# 	& frequency ratio, $\frac{\nu_{ref}}{\nu_{ion}}$ & \qec{} or $E_{ex}$ (keV) 	& mass excess (keV) \\
\\[-3mm]
\hline
\\[-3mm]
{\bf \nuc{46}{V}}		& {\bf \nuc{46}{Ti}}	& {\bf 40}	& {\bf 1.0001647674(71)}	& {\bf 7052.72(31)}
& {\bf -37070.68(86)} \\
\hline
\\[-3mm]
{\bf \nuc{42}{Sc}}	& {\bf \nuc{42}{Ca}}	& {\bf 52}	& {\bf 1.0001644199(52)}	& {\bf 6426.14(22)}
&  {\bf -32120.93(32)}\\	
\nucm{42}{Sc}		& \nuc{42}{Ca}	& 29	& 1.0001801961(54)	& 7042.73(23)	&  -31504.34(33)\\
\nucm{42}{Sc}		& \nuc{42}{Sc}	& 23	& 1.0000157743(58)	&  616.62(24)	&  -31504.64(35)\\
\multicolumn{4}{l}{{\bf Final superallowed \nuc{42}{Sc}---\nuc{42}{Ca} \qec{} value}}& {\bf 6426.13(21)} \\
\hline
\\[-3mm]
{\bf \nucm{26}{Al}}	& {\bf \nuc{26}{Mg}}	& {\bf 22}	& {\bf 1.0001748934(64)}	& {\bf 4232.79(15)}
&  {\bf -11981.79(16)}\\	
\nuc{26}{Al}		& \nuc{26}{Mg}	& 18	& 1.0001654660(64)	& 4004.63(15)	&  -12209.95(16)\\	
\nucm{26}{Al}		& \nuc{26}{Al}	& 18	& 1.0000094314(64)	&  228.30(16)	&  -11982.01(17)\\
\multicolumn{4}{l}{{\bf Final superallowed \nucm{26}{Al}---\nuc{26}{Mg} \qec{} value}}& {\bf 4232.83(13)} \\
\hline
\\[-3mm]
\nucm{26}{Al}		& \nuc{25}{Mg}	& 3	& 1.040075606(21)\phantom{0} & ~~~~~-	&  -11981.36(49) \\
\nuc{26}{Al}		& \nuc{25}{Mg}	& 4	& 1.040065775(19)\phantom{0} & ~~~~~-	&  -12210.15(45) \\		
\multicolumn{4}{l}{\nucm{26}{Al}---\nuc{26}{Al} using \nuc{25}{Mg} as reference }& 228.79(62) \\
\end{tabular}
\end{ruledtabular}
\end{center}
\end{table*}

There are three important conclusions we can draw from our \qec-value results. First, our result for
the superallowed \qec~value for \nuc{46}{V}, 7052.72(31)~keV, confirms the recent Savard
{\it et al.}~\cite{Sa05} measurement of 7052.90(40) keV, and disagrees with the previously accepted value
of 7050.71(89) keV, a survey result \cite{Ha05a} principally based on a 30-year-old ($^3$He,t)
$Q$-value measurement by Vonach {\it et al.} \cite{Vo77}.

Second, we can effectively rule out widespread systematic differences of more than $\sim$100~eV
between reaction-based $Q$-value measurements and those obtained with an on-line Penning trap.  In
their study of past measurements of (p,$\gamma$) and (n,$\gamma$) reaction $Q$ values near
\nuc{26}{Al}, Hardy {\it et al.} \cite{Ha06} derived a ``best'' reaction-based result for the mass
excess of \nuc{26}{Al} of -12210.27(11) keV.  By comparing reaction $Q$ values with much more precise {\it
off-line} Penning-trap measurements of stable nuclei in this same mass region, the authors cited
evidence for possible systematic effects in the former of $\sim$100 eV.  They then derived a second
mass excess for \nuc{26}{Al} of -12210.21(22) keV, a value that they state has been ``adjusted for
possible systematics.''  Our measurement of the \nuc{26}{Al} mass excess -- the first one made
with a Penning trap -- is -12209.95(16) keV and does not differ significantly from either of
the values presented by Hardy {\it et al.}; however, it certainly agrees more closely with
their systematics-adjusted value.  We cannot therefore
exclude systematic differences of up to $\sim$100 eV between reaction-based and {\it on-line}
trap measurements but anything significantly greater is ruled out.  This conclusion is further
supported by our \qec-value result for \nuc{42}{Sc}, 6426.13(21) keV, which agrees well with the
most precise previous result, 6425.84(17) keV, obtained from (p,$\gamma$) and (n,$\gamma$) reaction
$Q$ values \cite{Ha05a}.

This leads to our third conclusion, that no significant shift in the value of $V_{ud}$ should be
anticipated as more and more on-line Penning-trap measurements of the superallowed \qec-values
become available.  With our Penning-trap results for the \qec-values of \nucm{26}{Al} and
\nuc{42}{Sc} in good agreement with the previously accepted values \cite{Ha05a} and no evidence
now of significant systematic differences between reaction and Penning-trap measurements, it can
reasonably be concluded that \nuc{46}{V} was an anomalous case, for which only a single dominant
measurement had previously been available \cite{Vo77}, a measurement that appears simply to have 
been wrong.  For all other ``well known" superallowed transitions, several precise reaction-based
measurements already exist and new Penning-trap \qec-value measurements, when they appear, can safely be
averaged on an equal footing with those previous results.  To date, on-line Penning-trap results are
being quoted with uncertainties comparable to the best of the earlier measurements, so no large
changes should be expected in the resultant averages.

Although our result for \nuc{46}{V} confirms that there is a small anomaly in the $\F t$ value for
this transition \cite{Sa05}, if we incorporate our three new \qec-values and the one from ref.~\cite{Sa05}
into the 2005 survey data \cite{Ha05a} (and include improved radiative corrections \cite{Ma06}) we
find $V_{ud}$ = 0.9737(3), only marginally changed -- and slightly improved -- from the value 0.9738(4)
quoted in the survey.

The work was supported by the EU under contract numbers 506065 and HPRI-CT-2001-50034 and
by the Academy of Finland under the COE Programme 
2000--2005 (Project No. 44875).
JCH was supported by the U. S. Dept. of Energy under Grant DE-FG03-93ER40773 and by the Robert A. Welch
Foundation. 
AJ and HP are indebted to financial support from 
the Academy of Finland (Projects 46351 and 202256).

%\bibliography{PRLrefs_long}% Produces the bibliography via BibTeX.

\end{document}